\documentclass[pra,twocolumn,preprintnumbers,superscriptaddress]{revtex4}
\usepackage[utf8]{inputenc}
\usepackage{graphicx}
\usepackage{color}
\usepackage{dcolumn}
\usepackage{bm}

\usepackage{graphics}
\usepackage{amssymb}
\usepackage{times}
\usepackage{subfig}
\usepackage{amsbsy}
\usepackage{amsmath}
\usepackage{amsfonts}
\usepackage{amsthm}
\usepackage{float}
\usepackage{textcomp}
\usepackage{amsmath}
\DeclareMathOperator{\Tr}{Tr}

\DeclareMathOperator{\Cov}{Cov}

\newcommand{\bq}{\begin{equation}}
\newcommand{\eq}{\end{equation}}

\newcommand{\bqn}{\begin{eqnarray}}
\newcommand{\eqn}{\end{eqnarray}}

\newcommand{\ket}[1]{| #1 \rangle}
\newcommand{\bra}[1]{\langle #1 |}

\newcommand{\ketbra}[2]{| #1 \rangle \langle #2 |}

\theoremstyle{plain}

\begin{document}

\title{Mutual Uncertainty, Conditional Uncertainty and Strong Sub-Additivity}
\author {Sk Sazim}
\email{sk.sazimsq49@gmail.com}
\affiliation{Harish-Chandra Research Institute, HBNI, 
Chhatnag Road, Jhunsi, Allahabad 211019, India.}

\author {Satyabrata Adhikari}
\affiliation{Department of Mathematics, BIT Mesra, Ranchi-835215, India.}

\author {Arun K Pati}
\affiliation{Harish-Chandra Research Institute, HBNI, 
Chhatnag Road, Jhunsi, Allahabad 211019, India.}

\author {Pankaj Agrawal}
\affiliation{Institute of Physics, HBNI, Sainik School Post,
Bhubaneswar-751005, Orissa, India.}
\date{\today}

\begin{abstract}
We introduce a new concept called as the mutual 
uncertainty between two observables in a given quantum state which enjoys similar features like the mutual 
information for two random variables. Further, we define the conditional uncertainty as well as 
conditional variance and 
show that conditioning on more observable reduces the uncertainty. Given three observables, 
we prove a `strong sub-additivity' relation for the conditional uncertainty under certain condition. 
As an application, we show that using the conditional variance one can detect bipartite higher dimensional 
entangled states. The efficacy of our detection method lies in the fact that it gives better detection 
criteria than most of the existing criteria based on geometry of the states. Interestingly, we find that 
for $N$-qubit product states, the mutual uncertainty is exactly equal to $N-\sqrt{N}$, and if it is other 
than this value, the state is entangled. 
We also show that using the mutual uncertainty between 
two observables, one can detect non-Gaussian steering where Reid's criteria fails to detect. 
Our results may open up a new direction of exploration in quantum theory 
and quantum information using
the mutual uncertainty, conditional uncertainty and the strong sub-additivity for multiple
observables.
\end{abstract}
\maketitle
\section{Introduction}
In quantum theory, Heisenberg's uncertainty relation \cite{whu} restricts the knowledge 
of physical
observables one can have about the quantum system. The Heisenberg-Robertson
uncertainty \cite{hpru, kenn, book1, uncert_review} relation suggests the 
impossibility of preparing an ensemble
where one can measure two non-commuting observables with infinite precisions. 
Later, Schr\"{o}dinger \cite{esu}
improved the lower bound of this uncertainty relation. In fact, Robertson and 
Schr\"{o}dinger formulated mathematically the uncertainty relation for any two 
observables. 
Recently, the stronger uncertainty 
relations have been proved which go 
beyond the Robertson-Schr\"{o}dinger uncertainty relation 
\cite{str-ur1} and this has strengthened the notion of 
incompatible observables in quantum theory \cite{str-ur2, str-ur3, str-ur4, str-ur5, str-ur6, str-ur7}.
%
 
Shannon introduced entropy as a measure of information contained in a classical 
random variable \cite{shannon-ent}. The introduction of
entropy paved a path for a new field ``Classical Information Science'' \cite{info}. 
Later, von Neumann extended the idea of
entropy to the quantum domain where one replaces the probability distribution of 
random variables with the density operators for the states of  quantum systems.
Undoubtedly, entropy is an important quantity
in quantum information science  \cite{qinfo, wilde-book}. 
As entropy measures lack of information about the preparation of a
system, one can also express uncertainty relations in terms of entropies \cite{entro-unce, rev-entro}. 
However, in the quantum world, variance of 
an observable is also a measure of lack of information about the state preparation \cite{huang}.
Therefore, it may be natural to ask if using the variance as uncertainty measure, one can define analogous 
quantities such as the mutual information, the conditional entropy and the notion of strong sub-additivity.

Once we define these quantities, one immediate question is: Do they provide new insights about 
the quantum systems. The answer to this is in affirmative. 
For example, the mutual information, is the corner stone in defining many 
important aspects in information theory, like, unveiling correlations, channel capacities etc in quantum 
information science \cite{qinfo, wilde-book}. The conditional entropy is also inevitably an important quantity which is relevant in 
quantum communication as well as quantum computation \cite{qinfo, wilde-book}. While these analogies 
are very tempting to address for quantum uncertainty related quantities, there is a major departure between 
these two notions. The uncertainty is a function of both a quantum state and an observable whereas 
the notion of entropy depends on either of the two \cite{qinfo, wilde-book}. Moreover, while the uncertainty captures only the 
second moment, the entropy contains all the possible moments.

In this paper, we introduce the notion of mutual uncertainty, conditional uncertainty and 
strong sub-additivity on the basis of quantum uncertainties expressed in terms of standard 
deviations and variances. Interestingly,
we find that the standard deviation (quantum uncertainty) behaves in many ways like entropy. 
For example, we find that a chain rule for the sum uncertainty holds. Due to this fact one can easily define 
many important quantities like conditional mutual uncertainty as well.
Another important aspect of this formalism is that one 
can have a  version of `strong subadditivity (SSA)' for quantum uncertainties which may have implications 
in quantum information and this may be of independent interest. 
Also, we prove 
the strong subadditivity for more than three observables using the mutual uncertainty.

Then, we address the physical implication of all these quantities introduced here. 
As illustrations, we consider two important aspects in quantum information science -- 
detecting entanglement \cite{entrev, entdetrev,corr-rev} as well as quantum steering \cite{steer1}. 
We find that using the conditional variance, we can detect entanglement of higher dimensional bipartite 
mixed state. The method we present here is stronger than the criteria found by Vicente \cite{high_ent1}. 
Moreover, we find that for $N$-qubit product states, the mutual uncertainty is exactly equal to 
$N-\sqrt{N}$. This provides a sufficient condition to detect $N$-qubit entanglement. 
The other important finding is that 
we derive a steering criteria based on the mutual uncertainty. This criteria is as powerful as 
Reid's steering criteria \cite{MDsteer} for two qubits, 
and overpowers it when we consider non-Gaussian bipartite states. 
These results show the efficacy of our formalism. In fact, from the perspective of experimental realizations, 
our formalism might be one step ahead of the usual entropic formalism because, variances are easy to measure 
experimentally compared to entropic quantities which cannot be measured directly. 
%

The paper is organized as follows. 
In the next section, we discuss the sum uncertainty relation. Then, we define mutual 
uncertainty, conditional uncertainty and 
derive some important identities and inequalities like the chain rule, the strong subadditivity 
of uncertainties in section-III. 
In section-IV, we study the physical implication of these quantities, namely, usefulness of 
the conditional variance in detecting entangled states 
and finding steerable states using the mutual uncertainty. We conclude in the last section.

\section{Setting the stage: Sum uncertainty relations}
Let us consider a set of
observables represented by Hermitian operators $\{A_i\}$, then the
uncertainty of $A_i$ in a given quantum state $\rho$ is defined as
the statistical variance ($\triangle^2$) or standard deviation
($\triangle$) of the corresponding observable,
i.e., $\triangle A_i^2=\langle A_i^2\rangle-\langle A_i\rangle^2$, 
where $\langle A_i\rangle=\Tr[\rho A_i]$ for the state $\rho$. This positive
quantity can only be zero if $\rho$ is an eigenstate of $A_i$,
representing the exact predictability of the measurement outcome.
Hence, a quantum state with zero uncertainty must be a
simultaneous eigenstate of all $A_i$'s. 
The ``sum uncertainty relation'' \cite{arun1}  tells us that the
sum of uncertainty of two observables is greater or equals to the
uncertainty of the sum of the observables on a quantum system. If
$A$ and $B$ are two general observables that represent some
physical quantities,
then one may ask: What is the relation between $\triangle (A+B)$, $\triangle A$, 
and $\triangle B$? The following theorem answers this.\\
\textbf{Theorem.1} \cite{arun1}: {\it Quantum fluctuation in the sum of any two
observables is always less than or equal to the sum of their
individual fluctuations, i.e., $\triangle (A+B) \leq \triangle A +\triangle B$.}

The theorem was proved for pure states only but one can easily
extend the result for the arbitrary mixed states by employing the {\it
purification of the mixed states} in higher
dimensional Hilbert space. The physical meaning of the sum
uncertainty relation is that \textit{if we have an ensemble of
quantum systems then the ignorance in totality is always less than
the sum of the individual ignorances}. In the case of two observables,
if we prepare a large number of quantum systems in the state
$\rho$, and then perform the measurement of $A$ on some of those
systems and $B$ on some others, then the standard deviations in
$A$ plus $B$ will be more than the standard deviation in the
measurement of $(A+B)$ on those systems. Hence, it is always
advisable to go for the ‘joint measurement’ if we want to
minimize the error. Another aspect of this theorem is that it is
similar in spirit to the subadditivity of the von Neumann entropy,
i.e., $S(\rho_{12})\leq S(\rho_1)+S(\rho_2)$, where $\rho_{12}$ is
a two particle density operator and $\rho_2$=$\Tr_1(\rho_{12})$ is the
reduced density for subsystem $2$. Noticing this resemblance of
quantum entropy and standard deviation measure of uncertainty, it
is tempting to see if we can unravel some other features. Before
doing that we will first summarize the properties of the
uncertainty (captured by standard deviation) \cite{arun1}.

\noindent\textit{Properties of $\triangle(\cdot)$}: (i) $\triangle A_i\geq 0$ for $\{A_i\}$ in 
$\rho$, 
(ii) It is
convex in nature, i.e.,
$\triangle(\sum_ip_iA_i)\leq\sum_ip_i\triangle(A_i)$, with $0\leq p_{i}\leq 1$ 
and $\sum_{i}p_{i}=1$ and (iii)
one cannot decrease the uncertainty of an observable by mixing several
states $\rho=\sum_{\ell}\lambda_{\ell}\rho_{\ell}$, i.e.,
$\triangle(A)_{\rho}\geq
\sum_{\ell}\lambda_{\ell}\triangle(A)_{\rho_{\ell}}$, with
$\sum_{\ell}\lambda_{\ell}=1$. This is similar to the fact that entropy is also a
concave function of the density matrices, i.e.,
$S(\sum_{\ell}\lambda_{\ell}\rho_{\ell})\geq\sum_{\ell}\lambda_{\ell}S(\rho_{\ell})$. 

In fact, it is not difficult to see that if we have more than
two observables (say three observables $A$, $B$, and $C$), then the
sum uncertainty relation will read as
 $\triangle (A + B + C)\leq \triangle A +\triangle B +\triangle C$.
In general, for observables $\{A_i\}$, we will have
the sum uncertainty relation as
$\triangle (\sum_{i} A_i)\leq \sum_{i} \triangle A_i$ \cite{arun1}.

\section{Mutual uncertainty}
For any two observable A and B, the
mutual uncertainty in the quantum state $\rho$ is defined as
\begin{equation}
M (A:B):=\triangle A +\triangle B-\triangle (A + B).
\end{equation}
We name $M (A:B)$ as the \textit{mutual uncertainty} in the same
spirit as that of the mutual information. (The mutual information for a 
bipartite state $\rho_{12}$ is defined as
$I(\rho_{12})=S(\rho_{1})+S(\rho_{2})-S(\rho_{12})$.) 
The quantity $M (A:B)$ captures how much overlap two observables can have in a given 
quantum state.\\
\noindent {\it Properties of $M (A:B)$}: (i) $M (A:B) \geq 0$, (ii) it is symmetric 
in $A$ and $B$, i.e., $M (A:B)=M (B:A)$, and
(iii) $M (A:A)=0$. Note that $I(\rho_{12})$ also satisfies similar properties.\\ 
The above definition of mutual
uncertainty can be generalized for $n$ number of observables. Thus, given a set of 
observables $\{A_i; i=1,2,...,n\}$, we have 
\begin{eqnarray}
 M (A_1:A_2:\cdots:A_n):=\sum_{i=1}^n \triangle A_i -\triangle (\sum_{i=1}^nA_i).
\end{eqnarray}
The above relation is analogous to the mutual information for $n$-particle 
quantum state $\rho_{12..n}$ which is 
defined as $I(\rho_{12..n})=\sum_{i=1}^n S(\rho_i)-S(\rho_{12...n})$ \cite{note2}.

Note that all the observables may not have same physical dimension 
but one can make them of same 
dimension by multiplying them with proper dimensional
quantities. Although in this article we have omitted this
possibility by considering dimensionless observables.

\subsection{Conditional uncertainty and chain rule for uncertainties} 
We define a new quantity called the conditional uncertainty 
(similar to the conditional entropy
$S(\rho_{1|2})=S(\rho_{12})-S(\rho_{2})$) as
\begin{eqnarray}
 \triangle (A|B):&=&\triangle (A + B)-\triangle B.
\end{eqnarray}
This suggest that how much uncertainty in $(A+B)$ remains after we
remove the uncertainty in $B$.\\
\noindent {\it Properties of $\triangle (A|B)$}: (i)$\triangle
(A|B)\leq \triangle A$, i.e., conditioning on more observables 
reduces the uncertainty. (ii) $\triangle (A|B) \geq 0$ but can be
negative if $\triangle (A + B)<\triangle B$ or vice versa, (iii)
$\triangle (A|A)=\triangle A$.\\
By noting that $\triangle (A|B)=\triangle A-M(A:B)$ and $M(A:B)\geq0$, we have property (i). 
A simple example will illustrate the property (ii) \cite{note-ve}.

Now we will derive some useful results using the mutual uncertainty and the conditional uncertainty.  \\
\textbf{Theorem.2} {\it Chain rule for the sum uncertainty holds, i.e., 
$\triangle \left(\displaystyle\sum_{i=1}^{n}A_i\right)=\displaystyle\sum_{i=1}^{n} 
\triangle (A_i|A_{i-1}+\cdot\cdot\cdot+A_1)$.}\\
\begin{proof}
For three observables, the chain rule reads as
\begin{eqnarray}
 \triangle (A+B+C)
 = \triangle A+\triangle (B|A)+\triangle (C|A+B).\nonumber
\end{eqnarray}
Now consider 
\begin{eqnarray}
 RHS &=&\triangle A+\triangle (B|A)+\triangle (C|A+B)\nonumber\\
 &=& \triangle A+\triangle (B|A)+\triangle (C+A+B)-\triangle (A+B)\nonumber\\
 &=&  \triangle (C+A+B)=LHS.\nonumber
\end{eqnarray}
Similarly, one can prove by mathematical induction, that the theorem
holds for all positive integer $n$. 
\end{proof}

This tells us that
the sum uncertainty of two observable is equal to the uncertainty
of one observable plus the conditional uncertainty of the other
observables, i.e., $\triangle(A+B)=\triangle(A)+\triangle(B|A)$
which is similar to the entropy of the joint random variables
or
the bipartite systems. We can also define the following quantity as well.\\
\emph{Conditional mutual uncertainty}.-- We define another
quantity which we call the conditional mutual uncertainty in the same spirit of
the conditional mutual information. This is defined as 
$M (A:B|C):=\triangle (A|C)+ \triangle (B|C)-\triangle (A + B|C)$, 
which can be simplified as
\begin{eqnarray}
 M (A:B|C)=\triangle (B|C)-\triangle (B|C+A),
\end{eqnarray}
using the chain rule for the mutual uncertainty.

\subsection{Strong sub-additivity like relations}
The strong
sub-additivity of entropy is an important result in
information science. It gives a fundamental limitation to the
distribution of entropy in a composite system \cite{araki-lieb, lieb-ruskai}. In
classical case it implies the non-negativity of the mutual
information. For the relative entropy 
based quantum mutual information,
$I(\rho_{12..n})=S(\rho_{12..n}||\otimes_{i=1}^n\sigma_i)$
\cite{m-mutual-rela}, the strong sub-additivity of entropy guarantees the positivity \cite{posi-mu-m} but not for
the other versions of mutual information \cite{posi-mu-m1}. In a broad sense,
the strong sub-additivity of entropy implies that \textit{the conditioning will not increase the entropy}, i.e.,
$S(\rho_{1|23})\leq S(\rho_{1|2})$. Moreover,
beyond three particle systems we do not know the actual form of
strong sub-additivity of quantum entropy.\\
Here, we will prove a strong sub-additivity like 
relation concerning the uncertainties for multiple observables in a given quantum state.\\
\textbf{Theorem.3} {\it If $M(B:C)=0$, then $\triangle (A|B+C)\leq \triangle (A|B)$, i.e., 
conditioning on more observables reduces the uncertainty.}\\
\begin{proof} Lets start with the sum uncertainty relation, i.e., 
\bqn
\triangle (A+B+C)&\leq& \triangle (A+ B)+\triangle C\nonumber\\
\triangle (A+B+C)-\triangle (B+C)&\leq& \triangle (A+ B)-\triangle B
+\triangle B\nonumber\\&&+\triangle C -\triangle (B+C)\nonumber\\
\triangle (A|B+C)&\leq& \triangle (A|B)+M(B:C)\nonumber
\eqn
Hence, the proof. 
\end{proof}

The above relation can be understood as the ``Strong Sub-Additivity'' of uncertainty. 
The strong sub-additivity relation for uncertainty also ensures that the mutual uncertainty 
is always positive. For arbitrary number of
observables, the strong sub-additivity relation says that if $M(A_2+\cdot\cdot\cdot+A_{n-1}:A_n)=0$, 
then $\triangle
(A_1|A_2+\cdot\cdot\cdot+A_n)\leq \triangle
(A_1|A_2+\cdot\cdot\cdot+A_{n-1})$.

Next, we will prove two important relations concerning the mutual uncertainty.\\
\textbf{Inequality.1} Discarding the observable, 
one cannot increase the mutual uncertainty, i.e., $M(A: B)\leq M(A: B + C)$.\\
\begin{proof}
To prove this, let us start with the quantity $M(A:B+C)$.
\begin{eqnarray}
M (A:B+C)= \triangle A+\triangle (B+C)-\triangle (A+B+C)\nonumber\\
\triangle (A+B+C)=\triangle A+\triangle (B+C)-M
(A:B+C)\nonumber\\ \leq \triangle A+\triangle B+\triangle C -M
(A:B+C)\nonumber\\\leq  M (A:B)+\triangle (A+B)+\triangle C -M
(A:B+C)\nonumber\\ \leq (\triangle A+\triangle B+\triangle C)-(M
(A:B+C)-M (A:B)). \label{discardingnotincreasemu}\nonumber
\end{eqnarray}
Using the sum uncertainty relation for three observables 
\begin{eqnarray}
\triangle (A+B+C)\leq \triangle A+\triangle B+\triangle C\nonumber
\label{sumuncertaintythree}
\end{eqnarray}
and Eq.(\ref{discardingnotincreasemu}), we get
\begin{eqnarray}
M (A:B+C)-M (A:B))\geq 0.\nonumber
\label{discardingnotincreasemu1}
\end{eqnarray}
Hence the proof. 
\end{proof}

This is another form of strong sub-additivity in terms of mutual uncertainty. Interestingly, 
mutual information also satisfies $I(\rho_{12})\leq I(\rho_{1(23)})$ \cite{info}. Similarly, there 
is another total correlation measure, called as the entanglement of purification \cite{purif}, that 
satisfies $E(\rho_{12})\leq E(\rho_{1(23)})$ \cite{puriarn}. These observations provide added motivation 
to explore these new quantities in a greater details. 

All these inequalities resemble with the well known inequalities
concerning the entropy which are the corner
stone of quantum information science. However, we note that these similarities are structural, actual 
interpretations of these inequalities might be completely different.

\emph{Conditional variance}.-- Here, we define the conditional variance
(similar to the conditional entropy) as 
\begin{eqnarray}
 \triangle (A|B)^2:=\triangle(A+B)^2-\triangle{B}^2.
\end{eqnarray}
This quantity is equivalent to $\triangle A^2+2\Cov(A,B)$, where 
$\Cov(A,B)=\frac{1}{2}\Tr[\rho(AB+BA)]-\Tr[\rho A]\Tr[\rho B]$ is 
the covariance of $A$ and $B$. It
says that if the covariance is nonzero then the uncertainty in $A$ may increase or decrease  
due to the knowledge of the
uncertainty of $B$ as covariance can take both positive (correlation) and negative
(anti-correlation) values. This is in some sense different
from the conditional uncertainty. 

\section{Physical implications}
In this section, we will focus on some applications
of the quantities we introduced in the main text, eg., the 
mutual uncertainty, the conditional uncertainty and the conditional variance. We will study these
quantities for discrete systems such as the qubit-systems as well as higher dimensional systems, and 
the continuous variable systems also.

\subsection{Detection of entangled states}
Entanglement is a
crucial resource for many quantum information protocols (e.g., see
\cite{entrev}). Hence, detection and quantification of
entanglement is an important task. Several ways to
detect entanglement have been proposed in the recent past \cite{entdetrev}.
In the literature, the uncertainty relations have been 
employed to detect entanglement where operators can be either
locally applied on the subsystems \cite{hofmann} or globally
applied on the system as a whole \cite{guehne}. This motivates us
to ask the natural question here: Can we detect entanglement using
the conditional variance or other introduced quantities here? 
In the subsequent analysis, we answer this question in affirmative. 

There exists many elegant methods to detect entanglement using the local uncertainty relations 
\cite{hofmann,high_ent4, high_ent7, highD_Bloch_sep} or using geometry of quantum states \cite{cove1,maxE,local, high_ent1, 
high_ent2, high_ent3, high_ent5, high_ent6}. It is worthwhile to mention that 
using local uncertainty relations, one can detect more general form of entanglement, known as generalized 
entanglement which includes standard entanglement as special case \cite{sugg_1,sugg_2,sugg_3, sugg_4}.

In the following we use the conditional variance to derive a criteria which will detect the 
entanglement of two qudit mixed states. We find that the criteria based on the conditional variance is 
better than the existent criteria based on the geometry of the quantum states 
\cite{high_ent1, high_ent4, high_ent5, high_ent6, high_ent7}. We also consider $N$-qubit pure states 
and find a sufficient criteria of detecting its entanglement using the mutual uncertainty.

{\em Bloch representation of $N$-particle quantum systems and the condition for its separability}.-- 
To express quantum states in higher dimension geometrically, one need to understand the 
structure of $SU(d)$ group. It contains 
$d^2-1$ generators termed as $\sigma_{i}$, which form the basis of the Lie algebra with 
commutation and anti-commutation relations respectively
\bqn
[\sigma_{i},\sigma_{j}]=2i\sum_{k}f_{ijk}\sigma_{k},~~~~~~~ \nonumber\\
\{ \sigma_{i},\sigma_{j}\}=\frac{4}{d}\delta_{ij}
+2\sum_{k}d_{ijk}\sigma_{k}.\nonumber
\eqn
Here $f_{ijk}$ and $d_{ijk}$ are the anti-symmetric and symmetric structure constants.
All $\sigma_{i}$ are traceless 
Hermitian matrices which satisfy
$\sigma_{i}\sigma_{j}=\frac{2}{d}\delta_{ij}\mathbb{I}_d+\sum_{k}\left(if_{ijk}+ d_{ijk}\right)\sigma_{k}$.
For $d=2$, the symmetric structure constants, $d_{ijk}$ are ideally zero and the generators are 
well known Pauli matrices whereas for $d=3$, the generators are Gell-Mann matrices. 

Any arbitrary single particle quantum state in $d$-dimension can be expressed as 
$
\varrho=\frac{1}{d}\mathbb{I}_d+\frac{1}{2}\vec{r}.\vec{\sigma},
$ 
where $\mathbb{I}_{n}$ is the identity matrix of order $n$ and $|\vec{r}|^2\leq\frac{2(d-1)}{d}$. 
The density matrix, $\varrho$, is a Hermitian matrix with 
$\varrho\geq 0$, 
$\varrho\geq \varrho^2$ (equality holds when $\varrho$ is pure) and ${\rm Tr}[\varrho]=1$.

The $N$-qudit state can be expressed in the generalized Bloch vector representation as 
\begin{widetext}
\bqn
\rho=\frac{1}{d^N}\mathbb{I}_{d^N}+\frac{1}{2d^{N-1}}[\vec{\sigma}.\vec{r}\otimes\mathbb{I}_d^{\otimes N-1}+
\cdots + \mathbb{I}_d^{\otimes N-1}\otimes\vec{\sigma}.\vec{r_N}]
+\frac{1}{4d^{N-2}}\sum_{ij}[t_{ij0\cdots 0}\sigma_i\otimes\sigma_j\otimes\mathbb{I}_d^{\otimes N-2}
+\cdots\nonumber\\
+ t_{0\cdots 0ij}\mathbb{I}_d^{\otimes N-2}\otimes\sigma_i\otimes\sigma_j]+\cdots 
+\frac{1}{2^N}\sum_{i_1\cdots i_N} t_{i_1\cdots i_N} \sigma_{i_1}\otimes\cdots\otimes\sigma_{i_N},
~~~~~~~~~~~~~~~~~~~~~~~~~~~~~~~~~~
\label{N-qud}
\eqn
\end{widetext}
where $\vec{r_i}$ are the Bloch vector for the $i^{th}$ subsystem, 
$\{[t_{ij0\cdots 0}],\cdots,[t_{0\cdots0ij}]\}$ are 
pairwise correlation tensors, and $[t_{i_1\cdots i_N}]$ is the $N$-way correlation tensor. The are other type 
of correlation tensors, like, $3$-way, $4$-way, $\cdots$, $N-1$-way, which will not play a 
role in our analysis. For notational simplicity, we will call $T^{(k)}$ as $k$-way correlation tensor, where 
for example, $T^{(2)}$ forms a set $\{[t_{ij0\cdots 0}],\cdots,[t_{0\cdots0ij}]\}$ and so on.
The conditions required to approve the above matrix as a valid density matrix are -- 
$|\vec{r_i}|^2\leq\frac{2(d-1)}{d}$, $\rho\geq 0$, $\rho\geq \rho^2$ (equality holds when 
$\rho$ is pure) and ${\rm Tr}[\rho]=1$.

Now, we are ready to address the separability of the $N$-particle quantum state expressed in Eq.(\ref{N-qud}). 
This problem can easily be addressed by exploiting the Bloch-vector representation of the quantum systems as 
is shown in Refs.
\cite{multi_Bloch_sep1,multi_Bloch_sep2, multi_Bloch_sep3, multi_Bloch_sep4,multi_Bloch_sep5,multi_Bloch_sep6}. 
In order to describe the 
separability criteria, one can make use of the Ky-Fan norm \cite{kf_norm}. The Ky-Fan norm of 
a matrix, $X$, is defined as sum of the singular values ($\lambda_i$) of $X$, i.e., 
$||X||_{KF}:= \sum_i \lambda_i(X)$ $=\Tr[\sqrt{X^{\dagger}X}]$, where $\dagger$ denotes complex conjugation.  
In the Bloch-vector representation, for the state, $\rho$, if the reduced density matrix 
of a subsystem consisting of $k$ ($2\leq k \leq N$) out of $N$ parts is separable then 
$||T^{(k)}||_{KF}\leq \sqrt{(1/2^k) d^k(d-1)^k}$ \cite{multi_Bloch_sep1}. This is a set of conditions which 
leads to the hierarchy of entanglement structures \cite{entrev}. However, in this work, we are restricting 
our analysis for two qudit states and multi-qubit states. Note that for $N=2$, the 
separability condition is \cite{high_ent1}
\bq
||T||_{KF}\leq \frac{d(d-1)}{2}
\label{d_Sep}
\eq
and 
for $N$-qubit states ($d=2$), the separability conditions become $||T^{(k)}||_{KF}\leq 1$ 
\cite{multi_Bloch_sep1}.
\subsubsection{Detecting entanglement in higher dimensional bipartite quantum 
systems using conditional variance}

A bipartite quantum state of $d$-dimension is entangled when it cannot be expressed 
as $\rho=\sum_ip_i\rho_1^i\otimes\rho_2^i$. 
This means, for separable states, the correlation matrix can be expressed as  
$T=\sum_ip_i\vec{r_1}_i\vec{r_2}_i^{\intercal}$, where $p_i$ is the classical mixing parameter 
and $\intercal$ denotes the transposition. 
%
Here, we shed some light on the separability of the bipartite state using 
the quantities like the conditional variance and by exploiting the Bloch-vector representation of the state.  

Let $\mathcal{A}=\{\tilde{A}_i=\vec{a}_i.\vec{\sigma};i=1,...,d^2-1\}$ are a complete set of orthogonal 
observables such that ${\rm Tr}[\tilde{A}_i\tilde{A}_j]=2\delta_{ij}$. 
We can express these observables in a compact form like $\tilde{A}_i=\sum_j \Theta_{ij}\sigma_j$, where $\Theta$ 
$\in SO(d^2-1)$. Similarly, consider another such set of observables, 
$\mathcal{B}=\{\tilde{B}_i=\vec{b}_i.\vec{\sigma};i=1,...,d^2-1\}$, 
where $\vec{a}_i(\vec{b}_i)$ denotes the Bloch vector of the orthogonal operators 
$\tilde{A}_i(\tilde{B}_i)$ with unit norm. For observables like 
$A_i=\tilde{A}_i\otimes \mathbb{I}_d$ and $B_i=\mathbb{I}_d\otimes \tilde{B}_i$, the sum of all conditional variance is 
\bqn
\sum_i\triangle(A_i|B_i)^2=\sum_i\triangle (A_i+ B_i)^2-\sum_i\triangle B_i^2.
\eqn 
For two qudit separable states and the choice of above observables, we state the following theorem. 
\\
\textbf{Theorem-4} \textit{For two qudit separable states and the set of observables $\{A_i\}$ and $\{B_i\}$ 
described above, $\sum_i\triangle(A_i|B_i)^2 \geq 2(d-1)$. This criteria is equivalent to 
$||T||_{KF}\leq \frac{2(d-1)}{d}-\frac{1}{2}(|\vec{r_1}|-|\vec{r_2}|)^2$.}\\
\begin{proof}
 For the two qudit states, the sum of conditional variance can be expressed as
 \bqn
 \sum_i\triangle(A_i|B_i)^2 =\frac{2}{d}(d^2-1)+2\sum_i\vec{a}_i^{\intercal}T\vec{b}_i~~~~~~~~~~\nonumber\\
 -\sum_i(\vec{r_1}_i.\vec{a}_i+\vec{r_2}_i.\vec{b}_i)^2 + |\vec{r_2}|^2,\nonumber\\
 \leq  \frac{2}{d}(d^2-1)+2\sum_i\vec{a}_i^{\intercal}T\vec{b}_i 
 -(|\vec{r_1}|^2-2|\vec{r_1}||\vec{r_2}|).
 \label{condE1}
 \eqn
 While deriving the above relation, we have employed the fact that the symmetric structure constant 
 $d_{ijk}$ follows $\sum_{i=1}^{d^2-1} d_{iik}=0$, 
 $\forall k$.
 
 However, for two qudit separable states, the sum of conditional variance can directly be calculated as
 \bqn
 \sum_i\triangle(A_i|B_i)^2 
 =\frac{2}{d}(d^2-1)+2\sum_{i,j}p_j(\vec{r_1}_j.\vec{a}_i) (\vec{r_2}_j.\vec{b}_i)\nonumber\\
 -\sum_i\big[\sum_jp_j(\vec{r_1}_j.\vec{a}_i+\vec{r_2}_j.\vec{b}_i)\big]^2 + |\vec{r_2}|^2,\nonumber\\
 \geq  \frac{2}{d}(d^2-1)-\sum_jp_j(|\vec{r_1}_j|^2+|\vec{r_2}_j|^2)+ |\vec{r_2}|^2,\nonumber\\
 \geq 2(d-1),~~~~~~~~~~~~~~~~~~~~~~~~~~~~~~~~~~~~~~~~~~~~~~~~~~~~
 \label{condE2}
 \eqn
 where we used the relation, $2\sum_{j}p_j(\vec{r_1}_j.\vec{a}_i) (\vec{r_2}_j.\vec{b}_i)=\sum_j
 p_j[(\vec{r_1}_j.\vec{a}_i+\vec{r_2}_j.\vec{b}_i)^2-\{(\vec{r_1}_j.\vec{a}_i)^2+(\vec{r_2}_j.\vec{b}_i)^2\}]$. 
 The Eq.(\ref{condE2}) proves one part of the Theorem-4. 
 
 Now from Eqs.(\ref{condE1}, \ref{condE2}), one could 
 easily find that for two qudit separable states,
 \bq
 \sum_i\vec{a}_i^{\intercal} T\vec{b}_i \geq -\frac{2(d-1)}{d}+\frac{1}{2}(|\vec{r_1}|-|\vec{r_2}|)^2.
 \label{condF1}
 \eq
 The Eq.(\ref{condF1}) will be valid for any basis vectors $\vec{a}_i$ and $\vec{b}_i$. If we choose 
 $\vec{a}_i=\vec{u}_i$ and $\vec{b}_i=-\vec{v}_i$, where $\vec{u}_i$ and $\vec{v}_i$ 
 are left and right singular vectors of $T$ respectively, then the Eq.(\ref{condF}) can be casted as
 \bq
 ||T||_{KF} \leq \frac{2(d-1)}{d}-\frac{1}{2}(|\vec{r_1}|-|\vec{r_2}|)^2.
 \label{condF}
 \eq
Hence, the theorem is proved. 
\end{proof}

To show the efficacy of the new criteria, we have considered the following examples. \\
%
{\em Example.1}.-- Let us consider a two qubit state considered in canonical form,
$ \rho=\frac{1}{4}[\mathbb{I}_4+\frac{2}{5}(1-\alpha)\sigma_3\otimes\mathbb{I}_2
 -\frac{3}{5}(1-\alpha)\mathbb{I}_2\otimes\sigma_3-\alpha\sum_{i=1}^3\sigma_i\otimes\sigma_i],
$ 
which is entangled for $\alpha>\frac{1}{19(5\sqrt{6}-6)}\simeq 0.3288$ as predicted by Peres-Horodecki 
criteria \cite{peres-horo}. According to the new criteria, the above state is entangled when  
$\alpha>\frac{49}{74+5\sqrt{221}}\simeq 0.3303$ whereas the criteria in Eq.(\ref{d_Sep}) detects it for 
$\alpha>\frac{1}{3}$.
This example displays that the separability criteria derived in Eq.(\ref{condF}) is weaker than 
than the Peres-Horodecki criteria in $2\otimes 2$ dimension but it is stronger than the criteria in 
Eq.(\ref{d_Sep}). 

{\em Example.2}.-- Now, we consider the bound entangled state in $3\otimes 3$  
from Ref.\cite{TilesS}, i.e. 
$\rho=\frac{1}{4}[\mathbb{I}_9-\sum_i^4 \ketbra{\psi_i}{\psi_i}]$, where 
$\ket{\psi_0}=\ket{0}(\ket{0}-\ket{1})/\sqrt{2}$, $\ket{\psi_1}=(\ket{0}-\ket{1})\ket{2}/\sqrt{2}$, 
$\ket{\psi_2}=\ket{2}(\ket{1}-\ket{2})/\sqrt{2}$, $\ket{\psi_3}=(\ket{1}-\ket{2})\ket{0}/\sqrt{2}$ and 
$\ket{\psi_2}=(\ket{0}+\ket{1}+\ket{2})(\ket{0}+\ket{1}+\ket{2})/3$. For this state one readily finds 
that $||T||_{KF}\simeq 3.1603$, which violates both the conditions (\ref{condF}) and (\ref{d_Sep}). 
Hence, for this state, both the new criteria and the criteria in 
Eq.(\ref{d_Sep}) are able to detect its entanglement. Note that in 
this case, Peres-Horodecki criteria fails.
\subsubsection{Mutual uncertainty and the $N$-qubit pure states}
Before proceeding towards $N$-qubit pure states, we consider two qubit pure states.
Let us consider two observables $A=\vec{a}.\vec{\sigma}\otimes \mathbb{I}_2$ and 
$B=\mathbb{I}_2\otimes \vec{a}.\vec{\sigma}$ with $\vec{a}.\vec{r_1}=\vec{b}.\vec{r_2}=0$ and
$|\vec{a}|^2=|\vec{b}|^2=1$, where $\vec{\sigma}$ contains Pauli matrices only \cite{whytraceless}. 
Then for arbitrary pure two-qubit states the mutual uncertainty reads as 
$M(A:B)=2-\sqrt{2+2\vec{a^{\intercal}}T\vec{b}}$, where $T=[t_{ij}]$ is the correlation matrix. For pure 
product state, $T=\vec{r_1}\vec{r_2}^{\intercal}$, the mutual uncertainty turns out to be $M(A:B)=2-\sqrt{2}$. 
This result tells us
that if the mutual uncertainty for a given pure state is found to be other than
$2-\sqrt{2}\approx 0.586$, then the given pure state is entangled. This gives a sufficient 
condition for the detection of pure entangled state. Thus,
we can say that the mutual uncertainty between two observables can
detect pure entangled state. This may provide direct detection of pure entangled states in real experiment. 
Moreover, this is a state independent and observable independent universal value for the mutual uncertainty. 

There is another important aspect to this analysis for qubit systems. 
To show it, we consider arbitrary two qubit entangled state in Schmidt
decomposition form as $|\Psi\rangle=\sqrt{\lambda}|00\rangle+\sqrt{1-\lambda}|11\rangle$. 
The mutual uncertainty for the arbitrary pure two-qubit entangled state is given by 
\begin{eqnarray}
M(A:B)_{|\Psi\rangle}=2-\sqrt{2+2Ct},
\label{con}
\end{eqnarray}
where $C$ is the concurrence of $|\Psi\rangle$ \cite{conc} and $t=a_1b_1-a_2b_2$. Note that the concurrence 
of any arbitrary $|\Psi\rangle$ is defined as $C=|\langle\Psi|\sigma_2\otimes\sigma_2|\Psi^*\rangle|$, 
where $*$ complex conjugation. 
Interestingly, from the above relation one can see that by
measuring the mutual uncertainty between two observables, one can
directly infer the concurrence as $C = \frac{1}{2t}[2+ M(M-4)]$. Note that $t$ depends on the choice 
of observables.

The above analysis paves the way to extend the results for $N$-qubits. 
The mutual uncertainty expression for $N$-qubit state is
$
M(A_1:\cdots:A_N)=N-\sqrt{N+2\sum_{ij}\vec{a}_i^{\intercal} T^{(2)}\vec{a}_j},
$
where we have considered $A_i=\cdots\otimes\vec{a}_i.\vec{\sigma}\otimes\cdots$, 
with $i$ denotes the particular 
qubit and $\vec{a}_i.\vec{r}_i=0$, $|\vec{a}_i|^2=1$. (For example, 
$A_1=\vec{a}_1.\vec{\sigma}\otimes\mathbb{I}_2\otimes\mathbb{I}_2 \cdots$, 
$A_2=\mathbb{I}_2\otimes\vec{a}_2.\vec{\sigma}\otimes\mathbb{I}_2\otimes\cdots$ etc.) If the pure state 
is completely factorized then the bi-correlation matrices can be decomposed as 
$\{[t_{ij0\cdots}]=\vec{r}_1\vec{r}_2^{\intercal}, 
\cdots, [t_{\cdots0ij}]= \vec{r}_{N-1}\vec{r}_N^{\intercal}\}$. Hence, for 
genuine product states, 
\bq
M(A_1:\cdots:A_N)=N-\sqrt{N}.
\eq
Hence, we state the following proposition.\\
{\bf Proposition.1}.-- For pure $N$-qubit states with all pairwise correlation tensors of the form 
$T^{(2)}=\vec{r}_i\vec{r}_j^{\intercal}$ ($i\neq j$) and the set of $N$ observables $\{A_i\}$, 
the mutual uncertainty is $M(A_1:\cdots:A_N)=N-\sqrt{N}$, where $r_i$ is the Bloch 
vector of $i^{th}$ subsystem. 

Negation of the Proposition.1 for any pure $N$-qubit state sufficiently tells us that 
the state contains at least pairwise entanglement. Again, this provides a universal 
way to detect multiqubit entanglement.
\subsection{Detection of steerability of quantum states}
Quantum steering is a non-local phenomenon introduced by Schr\"{o}dinger \cite{steer1} 
while reinterpreting the EPR-paradox \cite{epr}.
The presence of entanglement between two subsystems in a bipartite state enables one to 
control the state of one subsystem by its entangled counter
part \cite{steer1,MDsteer}. Later, it was
mathematically formalized in Refs. \cite{detesteer, detesteer2}. 
Let Alice prepares an entangled state $\rho_{12}$ and sends one particle to Bob. Her 
job is to convince Bob that they are sharing non-local correlations (entanglement). Bob 
will believe such a claim if his state cannot be expressed by local hidden state model (LHS), i.e.,
$\tilde{\rho}_{1}^e=\sum_{\mu}p(\mu)\mathcal{P}(e|E,\mu)\rho_2^Q(\mu)$,
where $F=\{p(\mu),\rho_2^Q(\mu)\}$ is an ensemble prepared by Alice 
and $\mathcal{P}(e|E,\mu)$ is Alice's stochastic map. Here, $p(\mu)$ is the distribution of hidden variable 
$\mu$ with constraint 
$\sum_{\mu} p(\mu)=1$ and $E$ denotes all possible projective measurements 
for Alice. Conversely, if Bob cannot find such $F$ and $\mathcal{P}(e|E,\mu)$, then, he must admit that Alice 
can steer his system. Below, we present a strategy to detect quantum steering using the mutual uncertainty.

{\em Strategy}.--
To test whether a multiparticle state exhibits steering, one can devise an 
inequality based on the quantum properties of one of the particles and the inequality 
will be satisfied if the system has LHS model description. The violation of such inequality 
will be the signature of the steerability in the system. 

Here, we will devise such an inequality based on a simple property of the mutual uncertainty, i.e., 
$M(A:B)\geq 0$. We will employ the method used by Reid in Ref.\cite{MDsteer}. If two arbitrary observables, 
$A$ and $C$ has non-zero correlations, i.e., $\Cov(A,C)\neq 0$, then by knowing the measurement outcome 
of $C$ one can infer the value of $A$ which may reduce the error in the later measurement. Using this simple 
observation one can derive steering inequalities using different types of uncertainty relations 
\cite{und_steer0, und_steer1,und_steer3}. 

If Alice infers the measurement outcomes of $A$ performed by Bob, then the inferred uncertainty of 
$A$ is
\bq
\triangle_{\inf} A=\sqrt{\langle A-A_{est}(C)\rangle^2},
\eq
where $A_{est}(C)$ is the Alice's estimate using her measurement outcomes of $C$. 
In Ref.\cite{und_steer1}, it 
has been proved that the following inequality holds if we assume that Bob has LHS description 
\bqn
\triangle_{\inf} A +\triangle_{\inf} B \geq \triangle (A+B).\nonumber\\
\mbox{Hence,} \hspace{0.1cm} M_{\inf}(A:B)\geq 0, ~~~~~~~~~~~
\label{str_ineq}
\eqn
where $M_{\inf}(A:B)=\triangle_{\inf} A +\triangle_{\inf} B - \triangle (A+B)$, might be termed as the 
`inferred' mutual uncertainty. The Eq.(\ref{str_ineq}) is another type of steering inequality. \\
{\bf Proposition.2}.-- For any bipartite quantum state and any two observables, $A$ and $B$, 
if $M_{\inf}(A:B)< 0$, then the quantum state can demonstrate steering.

To demonstrate the power of the steering criteria in Proposition.2, we consider the following examples.\\
{\em Example-1}.--
In order to demonstrate our criteria in discrete systems, here, we will discuss 
the steerability of the Werner state,
\bq
\rho_{W}=p\ket{\Psi^-}\bra{\Psi^-}+\frac{1-p}{4}\mathbb{I}_4,
\eq
where $\ket{\Psi^-}=\frac{1}{\sqrt{2}}(\ket{01}-\ket{10})$ 
and $\mathbb{I}_4$ is the identity matrix of order $4$. The state $\rho_{W}$ is entangled 
for $p>\frac{1}{3}$, steerable for $p>\frac{1}{2}$ and Bell non-local for $p>\frac{1}{\sqrt{2}}$.

Let us consider two noncommuting observables, $A=\sigma_x/2$ and $B=\sigma_z/2$. In this case, 
the direct calculation shows that $M_{\inf}(A:B)=\sqrt{1-p^2}-1/\sqrt{2}$. Therefore, the Werner 
state will show steerability if $p> 1/\sqrt{2}$ for two measurement settings. However, there exist 
two measurement steering inequalities which are violated by Werner state for $p> 1/\sqrt{2}$ 
\cite{und_steer0, und_steer1,other_steer1}. Then the question is: 
what new features our new criteria entails. To show the power of our steering inequality, we will 
consider the following continuous variable systems. 

{\em Example-2}.-- 
We will consider the non-Gaussian state which can be created from a two-mode squeezed vacuum by 
subtracting a single photon from any of the two modes. The Wigner function of such a state in terms of 
the conjugate variables ($X_1, P_{X_1}$), ($X_2,P_{X_2}$) can be expressed as \cite{GSAgr},
\begin{widetext}
\bqn
W(X_1, P_{X_1}, X_2,P_{X_2}) &=& \frac{1}{\pi^{2}}\exp[2\sinh(2\alpha)(X_1X_2 - P_{X_1}P_{X_2}) 
 - \cosh(2\alpha)\sum_{i=1}^2 (X_i^{2}+P_{X_i}^{2})]
 [ - \sinh(2\alpha)\{(P_{X_1}-P_{X_2})^{2}\nonumber\\&-& (X_1-X_2)^{2}\} 
 +  \cosh(2\alpha)\{(P_{X_1} - P_{X_1})^{2}+ (X_1-X_2)^{2}\}-1],
\label{exmple2}
\eqn
\begin{figure}[H]
\centering
\includegraphics[height=8cm,width=13cm]{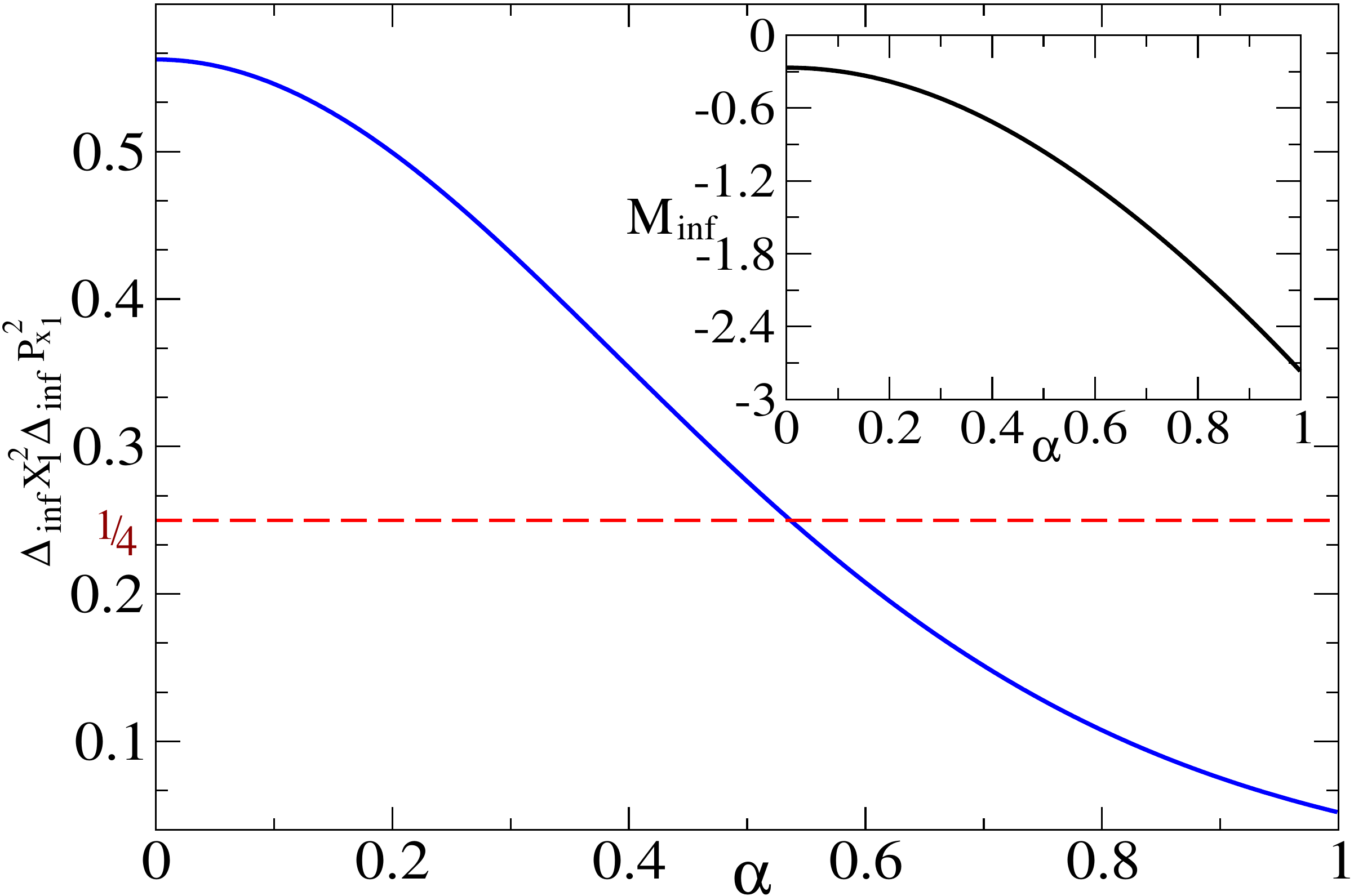}
\caption{(Color online) The graph shows the steerability of single photon subtracted squeezed 
vacuum state. The solid blue curve depicts the plot of the product of inferred uncertainties, 
$\triangle_{\inf} X_1^2\triangle_{\inf} P_{X_1}^2$ and the red dashed line represents the lower bound 
of Reid's inequality whereas the inset graph shows the plot of $M_{\inf} (X_1:P_{X_1})$ (solid black curve). 
It is clear that while the criteria based on the mutual uncertainty captures the 
steerability for any value of $\alpha$, the Reid criteria fails for 
$\alpha\leq \frac{1}{4}\cosh^{-1} (\frac{13}{3}) \approx 0.536$.}
\label{mu_steer_ng}
\end{figure}
\end{widetext}
where $\alpha$ is a squeezing parameter. Now, Alice will infer the conjugate 
observables ($X_1,P_{X_1}$) measured 
at Bob's by performing the observables ($X_2,P_{X_2}$) at her side. The inferred uncertainties can directly 
be calculated and hence, the inferred mutual uncertainty is  
\bq
M_{\inf} (X_1:P_{X_1})=\frac{\sqrt{3}}{2}\left(\frac{1}{ \eta_{-}}+\frac{1}{\eta_{+}}\right)
- (\eta_{+}+\eta_{-}),
\label{mu_cr}
\eq
where $\eta_{\pm}=\sqrt{\cosh(2\alpha) \pm \cosh(\alpha)\sinh(\alpha)}$. If $M_{\inf} (X_1:P_{X_1})<0$, then we can conclude 
that the state will demonstrate steering. To compare, we consider the Reid's criteria for steering which for 
our case is $\triangle_{\inf} X_1^2\triangle_{\inf} P_{X_1}^2\geq 1/4$ \cite{MDsteer}. For the state considered in 
Eq.(\ref{exmple2}), the right hand side of Reid's inequality comes out to be
\bq
\triangle_{\inf} X_1^2\triangle_{\inf} P_{X_1}^2=\frac{9}{2[3 \cosh(4\alpha)+ 5]}.
\label{rhsofR}
\eq
Now to draw comparison between two steering criteria,  we plot Eqs.(\ref{mu_cr} and \ref{rhsofR}). 
From the Fig.(\ref{mu_steer_ng}), we find that the steerability captured by the criteria based on 
mutual uncertainty is more than that of Reid's. More precisely, the criteria based on mutual uncertainty 
captures steerability for the whole range of $\alpha$ while the Reid's criteria fails for 
$\alpha\leq \frac{1}{4}\cosh^{-1} (\frac{13}{3}) \approx 0.536$.

\section{Discussions and conclusions}
We have introduced several new quantities called as the mutual 
uncertainty, the conditional uncertainty and the conditional variance
which may be useful in many ways to develop faithful notions in quantum information 
theory. In doing so, we have been able to prove many results similar to that
of entropic ones
such as the chain rule and the strong sub-additivity relations for the uncertainty. 
We have also shown that the conditional variance and the mutual uncertainty are 
useful to witness entanglement and quantum 
steering phenomenon. Specifically, as physical applications, we find that using 
the conditional variance, one can detect 
higher dimensional bipartite entangled states better than the criteria given in Ref.\cite{high_ent1}. Also, 
we find that the mutual uncertainty for $N$-qubit product states is exactly equal to $N-\sqrt{N}$, 
which provides a sufficient criteria to detect entanglement in multi-qubit pure states. 
Moreover, the steering criteria based on mutual 
uncertainty is able to detect non-Gaussian steering where Reid's criteria \cite{MDsteer} fails. 
In future, it may be interesting to see if these notions have other implications 
in quantum information science.

{\em Acknowledgement}.--
AKP gratefully acknowledges the local hospitality during his visits at IOP, Bhubaneswar 
for the period 2014-2016, where this work has been initiated. We would like to thank 
Sujit Choudhary for many stimulating feedback. We greatly acknowledge the effort of 
the anonymous referee, which enriched the presentation as well as the quality of our work.


\begin{thebibliography}{100}
\bibitem{note2} There exist no unique multiparticle generalization of quantum mutual information. We have 
considered the relative distance based one proposed in Ref.\cite{m-mutual-rela}.
\bibitem{note-ve} As an illustration of the fact that $\triangle (A|B) < 0$,
let us consider a Harmonic oscillator Hamiltonian
$H=\frac{1}{2}P^2+\frac{1}{2}X^2$. Let $A=\frac{1}{2}P^2$ and
$B=\frac{1}{2}X^2$. With this choice, one can see the quantity
$\triangle (A|B)$ can be negative. For example, if
$\ket{\psi}$ is an eigenstate of the Hamiltonian
then $\triangle (A + B)=0$ while $\triangle B$ is positive, and hence $\triangle
(A|B)$ can be negative.
\bibitem{whytraceless} The variance is invariant under the substraction of constant diagonal 
matrix, i.e., for observable $\mathcal{O}$, $\triangle (\mathcal{O}-k \mathbb{I})^2=\triangle \mathcal{O}^2$ 
with $k\in \mathcal{R}$ and $\mathbb{I}$ is the identity matrix. Therefore, without loss of 
generality one can consider traceless Hermitian operators as observables.
\end{thebibliography}
\end{document}